# An Features Extraction and Recognition Method for Underwater Acoustic Target Based on ATCNN

Gang Hu[1,2], Kejun Wang[1], Liangliang Liu[1]

[1] College of Automation, Harbin Engineering University, Harbin 150001, China

[2] College of Business, Anshan Normal University, Anshan 114007, China

Correspondence author: Kejun Wang (e-mail: wangkejun@hrbeu.edu.cn)

**Abstracts:** Facing the complex marine environment, it is extremely challenging to conduct underwater acoustic target recognition (UATR) using ship-radiated noise. Inspired by neural mechanism of auditory perception, this paper provides a new deep neural network trained by original underwater acoustic signals with depthwise separable convolution (DWS) and time-dilated convolution neural network, named auditory perception inspired time-dilated convolution neural network (ATCNN), and then implements detection and classification for underwater acoustic signals. The proposed ATCNN model consists of learnable features extractor and integration layer inspired by auditory perception, and time-dilated convolution inspired by language model. This paper decomposes original time-domain ship-radiated noise signals into different frequency components with depthwise separable convolution filter, and then extracts signal features based on auditory perception. The deep features are integrated on integration layer. The time-dilated convolution is used for long-term contextual modeling. As a result, like language model, intra-class and inter-class information can be fully used for UATR. For UATR task, the classification accuracy reaches 90.9%, which is the highest in contrast experiment. Experimental results show that ATCNN has great potential to improve the performance of UATR classification.

## 1. Introduction

Underwater acoustic target recognition (UATR) technology identifies target property [1] through analyzing ship-radiated noise received by sonar. Therefore, it is of great value for the economy and military. Because of the complex marine environment and application of acoustic stealth technology,

UATR has always been an internationally recognized problem. The traditional UATR method based on ship-radiated noise uses artificially designed features and shallow classifier to classify ships. Besides, traditional UATR work focuses on extracting features and developing non-linear classifier [2-7]. The features of artificially designed ship-radiated noise include waveform [8], spectrum [9], and wavelet [10], etc. These artificially designed features are weak in generalization because they depend on expert knowledge and prior knowledge. The shallow classifiers such as support vector machine (SVM) [11] and shallow neural network classifier [12] are weak in generalization and fitting when processing massive complex samples. Generally, the classifier design and feature extraction are conducted independently, consequently, designed features may be not applicable to the classification task. For example, the auditory filter set that is designed based on perception evidence in the classification model based on auditory features generally focuses only on signal properties rather than classification purpose [13].

Deep learning is an important branch of AI machine learning. It grows rapidly since Professor Hilton et al. proposed the concept of deep belief network (DBN) and related high-efficient learning algorithm in 2006 [14,15]. Wikipedia defines deep learning as "a set of complex data modeling algorithms with multi-layer nonlinear transformation". The neuroscientists found that human auditory system was unique in acoustic recognition. This is because human brain has strong capacity in perception, inference, induction, and learning. Inspired by neural structure of human and information processing mechanism of the brain, researchers proposed deep neural networks (DNNs) that processed information and decision-making in a way similar to human brain [16]. With deep learning technology, it is possible to conduct mathematical modeling for original signals and predict objectives in a complete deep learning model. And people also believe that human auditory system works in a similar way. These findings show that time-domain acoustic signals of the auditory system can be decomposed into frequency components; different areas of the auditory system perceive information of different frequency components; the brain collects information of all areas for analysis and for classifying acoustic signals. In addition, the research on shaping of auditory cortex has demonstrated that adult brain can be reshaped in appropriate environment, and auditory experience can change functions and even structure of the auditory system [17]. The neuroscientists found that human auditory system was unique in acoustic recognition. This is because human brain has strong capacity in perception, inference, induction, and learning. Inspired by neural structure of human and information processing mechanism of the brain, researchers proposed deep neural networks (DNNs) that processed information and decision-making in a way similar to human brain.

In addition, the language model method has been widely used in natural language model [17,18,19]. This method is used in the RNN-based sequence-to-sequence model, and the first input of decoder RNN can be adjusted with prediction of the model classifier. K. Drossos et al. used the language model for detecting acoustic events [20] and achieved good effects. The language model method enables RNN to adjust input of RNN with previous prediction of the classifier, process long messages, and implement contextual modeling for class activities, thus learning longer intra-class and inter-class time model. When being used for UATR, the language model method can improve performance through modeling for these class-dependency relationships [20].

Inspired by auditory perception and language model, this paper proposes an end-to-end ATCNN that is used for UATR. The proposed ATCNN consists of a series of depthwise separable convolution subnets, integration layers, and time-dilated convolutions. In the depthwise separable convolution subnet, the depthwise separable convolution filter with changed width of convolution kernel is used to decompose original time-domain ship-radiated noise signals into different frequency components, and extract signal features based on auditory perception. Moreover, deep features are integrated on the integration layer. The time-dilated convolution is used for long-term contextual modeling. As a result, like language model, intra-class and inter-class information can be fully used for UATR. The proposed ATCNN here is matched with information integration on auditory cortex and the function of acoustic signal recognition. Therefore, this paper provides a method that can adjust time-dilated convolution with predictions of the classifier. The dataset of measured underwater target radiated noise, and indicators based on typical calculation and per frame calculation (i.e.F1 value and accuracy) are used to evaluate performance of our method.

The main contributions of this paper include:

1. In the model, this paper provides a depthwise filter sub-network that consists of depthwise separable convolution filter to simulate the structure of extracting depthwise acoustic information in acoustic system.

2. Inspired by perception neural mechanism of frequency component, the complex frequency component of ship-radiated noise is decomposed, and a group of multi-scale filtering subnets of the depthwise separable convolution are used for modeling.

3. Inspired by shapeable neural mechanism, parameters of multi-scale filtering subnets of depthwise separable convolution are learnt from original time-domain ship-radiated noise signals.

4. Inspired by the language modeling mechanism, the time-dilated convolution is used (expanded dilated convolution is used only on time dimension) to learn intra-class and inter-class information for UATR.

5. According to experimental results, the proposed ATCNN model is effective for UATR. This method can effectively realize decomposition, modeling, and classification for ship-radiated noise signals, and has achieved better classification results than existing methods.

This paper is organized as below: Section 2 introduces related work; Section 3 introduces the proposed method in detail; Section 4 describes the following evaluation process; Section 5 introduces and discusses experimental results; Section 6 discusses the conclusions.

## 2. Related work

All existing studies on passive UATR with deep learning now are still in preliminary stage and focus on theoretical exploration and small-scale experiments. Generally, application of deep learning shall be combined with big data. However, due to conditional limitation, it is usually impossible to collect enough data for model training, significantly limiting the performances of DNN. Nevertheless, urgent demands still promote continuous growth of deep learning in passive UATR. CNN is widely used in passive UATR because it is applicable to process original underwater acoustic signals and can obtain hidden correlation difficult to be found with common feature analysis methods to a certain extent [21, 22, 23, 24]. With internal feedback mechanism, RNN can process sequence signals. The audio signal is a typical sequence signal, which is provided with memory function by RNN through circumferential joints of internal neurons. Therefore, it can dynamically utilize correlation of acoustic signals on time dimension. There is a typical structure called as "Long Short Term Memory" (LSTM) in RNN [17]. There are studies that use LSTM network in passive UATR [25, 26, 27].

Among published papers about audio classification, CNN is replaced with DWS convolution [28, 29]. DWS convolution is a decomposition form of standard convolution, which decomposes a standard convolution into one convolution and one 1×1 convolution (called as "pointwise convolution) [30]. It firstly learns spatial information and then processes cross-channel mode [29, 31]. This convolutional decomposition for typical CNN results in smaller training parameters and memory. Moreover, computation complexity declines $K_o^{-1} + (K_h \cdot K_w)^{-1}$, where $K_h$ and $K_w$ are height and width of CNN kernel, and $K_o$ is output channel of CNN [28]. The dilated convolution is considered as a method of improving long-term leaning capacity of CNN [32]. In short, the kernel of dilated convolution is dilated and there is a distance between two elements. As a result, the kernel of dilated convolution can be used on elements of its input patch with interval N (dilation factor), increasing receptive field of the kernel instead of its parameters [32, 33, 34]. The kernel dilation can be used in any combination (for example, dilation on time dimension or feature dimension only) or all combinations of its dimensions. Y.Li et al. provided a method of using dilated convolution and RNN in audio classification task [35], which especially focused on exploring and learning long-term model. K.Drossos et al. provided an improved CRNN structure [28] which used DWS and dilated convolution with dilation on time dimension only, i.e. time-dilated convolution. With scattered wavelet and time-dilated convolution, this structure had 85% lesser parameters than CRNN, but achieved better performances in typical audio classification dataset. According to improvement for dilated convolution, these convolutions have similar functions with RNN and can be effectively used for long-term contextual modeling.

For human, acoustic perception and recognition are completed in auditory system, including auditory periphery and center [36]. The auditory nerve generates acoustic signals as below: first, the auditory nerve receives acoustic signals; then, acoustic frequency, strength and other information are transferred to auditory center through auditory nerve; finally, the information is integrated and recognized on auditory cortex. With development of neuroscience, the neural mechanism of auditory perception is disclosed. Researchers find that some structures of the deep auditory system have frequency decomposition capacity with different range and resolution, for example, cochlea [37, 38], auditory midbrain [39], primary auditory cortex, and secondary auditory cortex [40, 41, 42]. When analyzing frequency, the primary auditory cortex decomposes complex acoustic signals into different frequency components through nerve cells with multi-scale receptive field of frequency [40]. In addition. different frequency components in acoustic signals can activate different areas in the auditory system. The complex signals with multiple frequency components can activate more areas which are widely distributed on the primary and secondary auditory cortex

[41, 42]. Other researchers are devoting to studying the shaping of the brain. This means that the brain can adjust its structure and functions to meet learning demands [43]. The frequency perception related areas in structures such as auditory center, auditory cortex, auditory midbrain can adjust receptive field of frequency and the optimal frequency to complete learning task [44, 45]. These findings about auditory system indicate that the acoustic time-domain signals can be decomposed as per frequency component in the auditory system. The decomposition of frequency components can be explained as product filtering for acoustic frequency-domain signals. The product of frequency domain signal is equal to convolution of time-domain signal [46], therefore, the frequency-domain component can be rapidly achieved through parallel computation of time-domain convolution; different areas of the auditory system perceive different frequency components; the brain collects information of all areas for analysis and for classifying acoustic signals. In addition, studies on the shaping of auditory cortex have demonstrated that adult brain can be reshaped in appropriate environment. The auditory experience can change functions and even structure of the auditory system.

## 3. Auditory perception inspired time-dilated convolution neural network for UATR

### 3.1 The Architecture of ATCNN for UATR

Sonar staff can recognize their interested targets from complex marine environment depending on strong information processing capacity of the auditory system. Based on findings of neurosciences, this paper summarizes structure of extracting deep auditory information of the auditory perception and some neural mechanism for reference. The structure of deep acoustic information of auditory system is a multilayer system which mainly includes cochlea, auditory midbrain, auditory thalamus, and auditory cortex. Different frequency components can be decomposed in the auditory channel from cochlea, midbrain, to auditory cortex. The above neural mechanisms include frequency-component perception mechanism and shaping mechanism. For frequency-component perception mechanism, different areas such as cochlea, auditory midbrain, primary auditory cortex, and secondary auditory cortex can perceive decomposition of different frequency components, while decomposition with similar frequency component can activate relatively fixed areas. For shaping mechanism, the auditory system continuously adjusts auditory perception with the drive of different auditory simulation in different learning tasks or environment to meet learning demands. The shaping of auditory system exists throughout auditory perception. Inspired by auditory perception, ATCNN is used for ship-radiated noise modeling and ship classification. This model contains a series of DWS convolution subnets, integration layers, and time-dilated convolutions. Figure 1 shows the architecture of ATCNN.

The proposed model uses sequence $X \in \mathbb{R}^{T \times N}$ of vector $T$ as input, and each vector $T$ consists of original underwater acoustic time-domain data with length $N$. Besides, this model also uses learnable feature extractor that consists of DWS convolution subnet and time-dilated convolution as time mode recognizer. $X$ corresponds to label vector $Y = [y_1, \cdots, y_C]$ of C classes, where, $y_c \in \{0,1\}$ is $c^{th}$ class or not when $X$ is input. The model output is a vector $\hat{Y} = [\hat{y}_1, \cdots, \hat{y}_C]$, where $\hat{y}_c \in [0,1]$ is classification result predicted by the model for underwater acoustic target when $X$ is input. Inspired by the structure of extracting deep acoustic information of auditory system, this paper designs a series of DWS filter subnets. These subnets input original underwater acoustic data with the one-dimensional DWS CNN. In addition, DWS convolution subnets have realized frequency decomposition for input signal and extracted features of decomposed signals.

The feature integration layer integrates information of realizing multi-channel feature per time-domain with pointwise convolution. On this feature integration layer, all one-dimensional acoustic features output by DWS filter subnets at time $T$ are integrated and analyzed comprehensively. The integrated acoustic feature is two-dimensional time mode feature at time $T$ and therefore can be used as input of time-dilated convolution layer. Besides, the two-dimensional frequency convolution layer is used to maintain local completeness and reduce frequency spectrum change of ship-radiated noise.

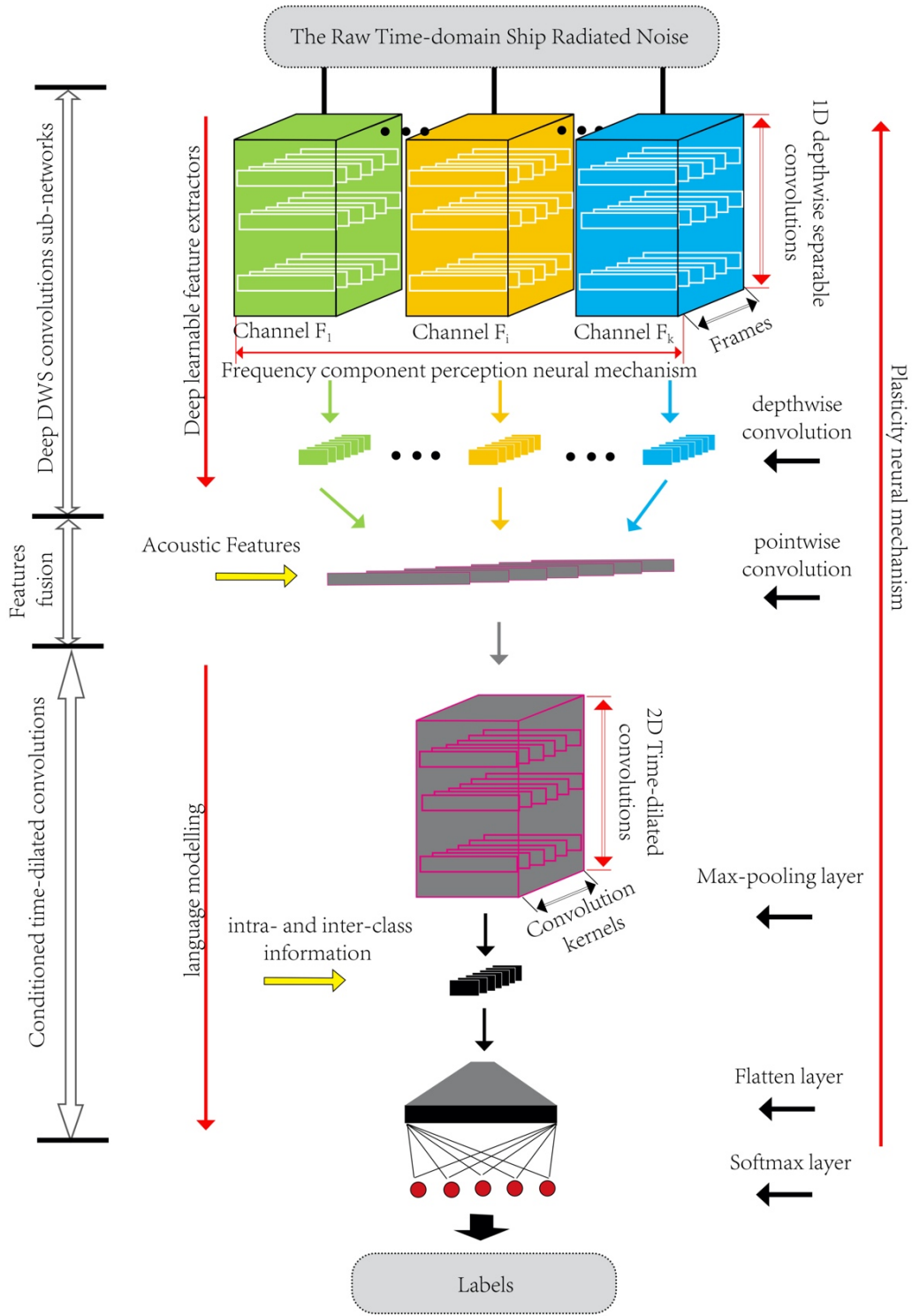

Figure 1. The architecture of ATCNN.

In order to utilize intra-class and inter-class activity mode on time-dilated convolution layer, this paper uses a time-dilated convolution like language modeling, and uses Softmax classifier to obtain a prediction probability of each ship in each sample. Moreover, this paper uses ship classification as target function, with drive of original ship-radiated noise, learns and optimizes the whole training process. This optimization mechanism reflects the shaping mechanism of the auditory system.

This model can realize decomposition, feature extraction and classification for ship-radiated noise and therefore can be used for UATR task.

**3.2 Learnable feature extractor inspired by auditory perception**

The cochlear response has been widely studied. Signals are coded with a group of kernels in cochlea. The kernel can be considered as an array of overlapped band-pass auditory filters. The center frequency of these filters increases from top to bottom of cochlea. In addition, their bandpass is smaller under low frequency [47, 48]. Because the energy of ship-radiated noise gathers at low frequency, this feature is applicable to describe ship-radiated noise.

CNN is an ANN signal convolution that conducts a series of convolutions for input. The operation in CNN is equivalent to time-domain convolution in traditional filter [49]. This paper designs multi-layer CNN on each depth filter subnetwork to realize filtering function. Therefore, it is defined as depthwise separable convolution filter. Through repeating above process layer-by-layer, the multi-layer CNN built can extract more abstract features from deep structure. However, deeper unit can be indirectly connected with all or most of signals. The receptive field of deep unit on depthwise filter subnet is larger than the shallow unit [50]. The parameters of depthwise separable convolution filter are randomly initialized and learnt from ship-radiated noise. With the drive of time-domain signals of ship-radiated noise, frequency decomposition of depthwise separable convolution filter can be learnt and adjusted. In addition, larger convolution kernel can contain longer wavelength, which means low frequency of the component, and vice versa. As a result, learnt filters are more applicable to UATR task.

In order to learn spatial information and cross-channel information, this paper uses two different kernels connected with two convolutions in series instead of convolution of a single kernel (i.e. output of the first convolution is input of the second convolution). This decomposition technology is called as DWS convolution which has been used in multiple image processing architectures, for example, Exception, GoogleLeNet, Inception and MobileNets model. Besides, it has been proved that DWS convolution can reduce parameter quantity and improve performance [51, 52, 53]. DWS convolution consists of depthwise convolution and pointwise convolution. This paper uses a filter for each input channel (input depth) with depthwise convolution. The pointwise convolution is a simple 1×1 convolution and used to create linear combination for output of depthwise layer. The learnable feature extractor consists of $L$ number of DWS convolution blocks. The $l$th DWS convolution block obtains output of the previous block as input, i.e. $H_{l-1} \in \mathbb{R}^{H^c_{l-1} \times H^h_{l-1} \times H^w_{l-1}}$, where $H^c_{l-1}$, $H^h_{l-1}$ and $H^w_{l-1}$ are channel number, height and width output by the $l$-1$^{th}$ DWS convolution block. $H_0 = X$ is input time-domain signal of the ship-radiated noise, $H^c_0 = 1$, $H^h_0 = 1$, $H^w_0 = N$. The output of the $l^{th}$ DWS convolution block is $H_l \in \mathbb{R}^{H^c_l \times H^h_l \times H^w_l}$, where $H^c_l$, $H^h_l$ and $H^w_l$ are channel number, height and width output by the $l^{th}$ DWS convolution block. Each DWS convolution block includes one DWS convolution operation, one normalization, one down-sampling, and one non-linear function.

The operation of DWS convolution includes two convolutions, one normalization process and one rectified linear unit (ReLU). The first convolution on the $l^{th}$ DWS convolution block uses $H^c_{l-1}$ number of kernels $K_l \in \mathbb{R}^{K^{dh}_l \times K^{dw}_l}$ and one bias deviation $a_l \in \mathbb{R}^{H^c_{l-1}}$ to learn spatial information input into $H_{l-1}$. $K^{dh}_l$ and $K^{dw}_l$ are height and width of the kernel $K_l$; and $s_l$ is stride of the convolution kernel $K_l$. The depthwise convolution with a filter in each input channel (input depth) can be written as:

$$D_l^{h^c_{l-1},d^h_l,d^w_l} = \left( H^{h^c_{l-1}}_{l-1} \otimes K^{h^c_{l-1}}_l \right)(d^h_l, d^w_l)$$

$$= \sum_{k^{dh}=1}^{K^{dh}} \sum_{k^{dw}=1}^{K^{dw}} H_{l-1}^{h^c_{l-1}, s_l \cdot d^h_l + k^{dh}, s_l \cdot d^w_l + k^{dw}} K_l^{h^c_{l-1}, k^{dh}, k^{dw}} + a_l^{h^c_{l-1}} \qquad (1)$$

where, $D_l \in \mathbb{R}^{H^c_{l-1} \times D^h_l \times D^w_l}$ is output of the first convolution on the $l^{th}$ DWS convolution block. $D^h_l$ and $D^w_l$ are height and width of $D_l$. Figure 2 shows operation process of the first convolution on DWS convolution.

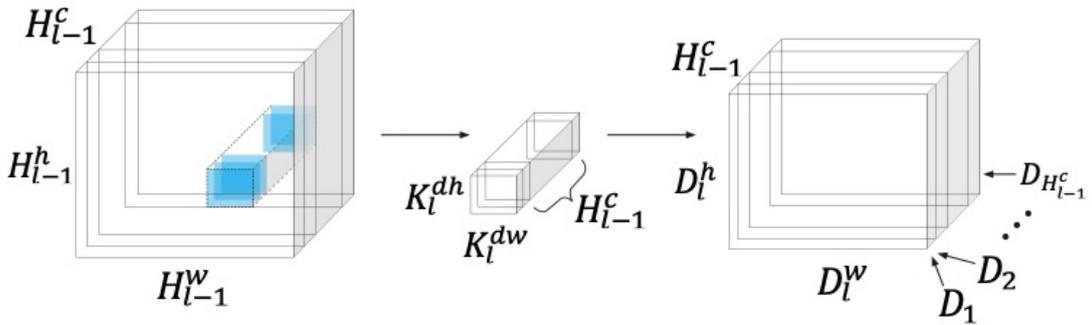

Figure 2. The first step of depthwise separable convolution. Learning spatial information, using $H^c_{l-1}$ different kernels $K_l$, applied to each $H_{l-1}$.

Next, $D_l$ is input into ReLU activation function after batch normalization (BN). The process is described as below:

$$D'_l = ReLU(BN(D_l)) \quad (2)$$

Where, ReLU is non-linear activation function of linear rectification; BN is batch normalization; $\boldsymbol{D'_l} \in \mathbb{R}^{H^c_{l-1}, D^h_l, D^w_l}$ is output of $\boldsymbol{ReLU}$. The second input of DWS convolution is $\boldsymbol{D'_l}$. In addition, $\boldsymbol{H^c_l}$ number of 1×1 convolution kernels $\boldsymbol{Z_l} \in \mathbb{R}^{H^c_{l-1}}$ and a bias vector $\boldsymbol{b_l} \in \mathbb{R}^{H^c_l}$ are used to learn cross-channel information, with process as below:

$$S_l^{h^c_l, d^h_l, d^w_l} = \left(D'^{h^c_{l-1}}_l \otimes Z^{h^c_l}_l\right)(d^h_l, d^w_l)$$

$$= \sum_{h^c_{l-1}=1}^{H^c_{l-1}} D'^{h^c_{l-1}, d^h_l, d^w_l}_l Z^{h^c_l, h^c_{l-1}}_l + b^{h^c_l}_l \quad (3)$$

Where, $S_l \in \mathbb{R}^{H^c_l, D^h_l, D^w_l}$ is output of the second convolution block on the $l$ the DWS convolution. Figure 3 shows operation process of the second convolution on DWS convolution.

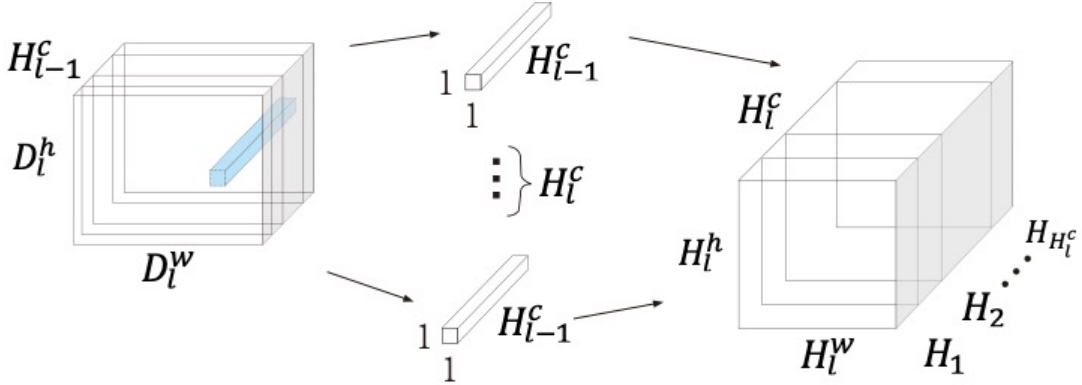

Figure 3. The second step of depthwise separable convolution. Learning cross-channel information using $\boldsymbol{H^c_l}$ different kernels $\boldsymbol{Z_l}$.

The output of DWS convolution block can be expressed as below:

$$H_l = ReLU(BN(S_l)) \quad (4)$$

The down-sampling is used on application feature dimension after each $H_l$ of DWS convolution block, for example, maximum pooling. In Equation (1) & (3), $O(H^c_{l-1} \cdot K^{dh}_l \cdot K^{dw}_l \cdot$

$D_l^h \cdot D_l^w + H_l^c \cdot H_{l-1}^c \cdot D_l^h \cdot D_l^w$) and $H_{l-1}^c \cdot K_l^{dh} \cdot K_l^{dw} + H_{l-1}^c \cdot H_l^c$ are computation complexity and parameter quantity (neglecting deviation). Therefore, computation complexity [30] and parameter quantity have declined $(H_l^c)^{-1} + (K_l^{dh} \cdot K_l^{dw})^{-1}$ times compared to operation of standard convolution with same functions. The final output of DWS convolution block $H_L \in \mathbb{R}^{H_L^c \times H_L^h \times H_L^w}$ is compressed into one-dimensional vector $H' \in \mathbb{R}^F$, where $H'$ is one-dimensional feature vector that is extracted by learnable feature extractor inspired by auditory perception; and F is length of the feature vector.

**3.3 Time-dilated convolution inspired by language model**

The dilated convolutions introduce a new parameter to convolution layer, called as "dilation rate". This parameter defines the gap between values during data processing with convolution kernel. The structure aims to provide larger receptive field with equivalent computing amount and without pooling (pooling layer may result in information loss). Because the dilated convolutions can cluster and learn multi-scale information, they are widely used in visual field of deep learning now and have achieved outstanding performances in target detection and image segmentation [32]. In addition, feature extractor obtains one-dimensional underwater acoustic feature vector $H' \in \mathbb{R}^F$ at continuous $T$ time, and then combines T number of $H'$ into two-dimensional matrix $I \in \mathbb{R}^{T \times F}$ on integration layer, where $T$ and $F$ respectively represent height and width of matrix, and $I$ is input of two-dimensional time-dilated convolution. This paper uses long-term mode of $I$ with the two-dimensional time-dilated convolution, and then a linear layer with softmax activation function. This layer can be used as classifier.

The time-dilated convolution network here used for long-term mode consists of $J$ number of time-dilated convolution blocks. The $j$th time dilated convolution block obtains output of the previous block as input, i.e. $O_{j-1} \in \mathbb{R}^{O_{j-1}^c \times O_{j-1}^h \times O_{j-1}^w}$, where, $O_{j-1}^c$, $O_{j-1}^h$ and $O_{j-1}^w$ are channel number, height and width output by the $j-1$th time-dilated convolution block. $O_0 = I$ is the feature matrix of the input ship-radiated noise, $O_0^c = 1$, $O_0^h = T$, $O_0^w = F$. The output of the $j$th time-dilated convolution block is $O_j \in \mathbb{R}^{O_j^c \times O_j^h \times O_j^w}$, where $O_j^c$, $O_j^h$ and $O_j^w$ are channel number, height and width of the $j$th time-dilated convolution block. The $j$th time-dilated convolution consists of $O_j^c$ kernels $K_j' \in \mathbb{R}^{O_{j-1}^c \times K_j'^h \times K_j'^w}$ and bias vector $b_j' \in \mathbb{R}^{O_j^c}$, where $K_l^{dh}$ and $K_l^{dw}$ are height and width of the kernel $K_j'$.

$$Q_j^{o_j^c, o_l^h, o_l^w} = \left(O_{j-1}^{o_{l-1}^c} \otimes K_j'^{o_{l-1}^c}\right)(o_l^h, o_l^w)$$
$$= \sum_{o_{j-1}^c=1}^{O_{j-1}^c} \sum_{k^{dh}=1}^{K^{dh}} \sum_{k^{dw}=1}^{K^{dw}} O_{j-1}^{o_j^c, o_l^h + \xi^h \cdot k^{dh}, o_l^w + k^{dw}} K_j'^{o_j^c, o_{j-1}^c, k^{dh}, k^{dw}} + b_j'^{o_j^c} \quad (5)$$

Where, $\xi^h$ is dilation rate of $K_j'$ at $K_j'^h$ dimension. It shall be noted that dilation is conducted only on time dimension in this paper. The dilation rate $\xi^h$ multiplying by ( ) is used for visiting index $k^{dh}$ of the element $O_{j-1}$. This allows to cluster contextual information in proportion at output of operation [32]. In fact, this means that the feature result calculated with time-dilated convolution is calculated from a larger area. Consequently, longer time context can be used to create recognition model. The process described in Eq. (5) is shown in Figure 4.

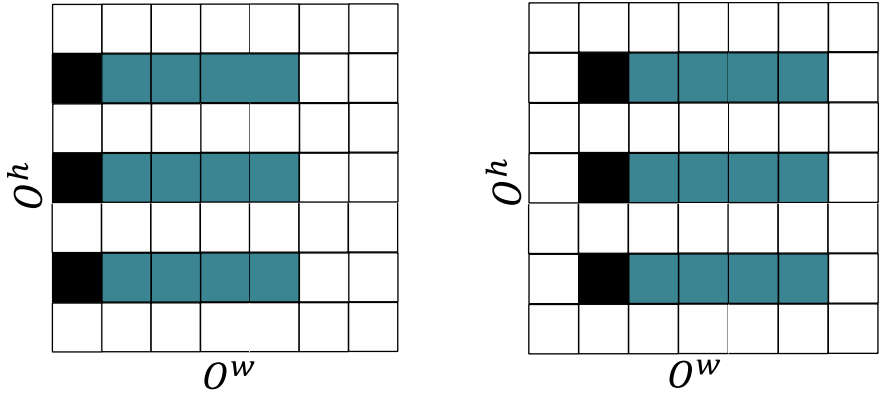

(a) Processing of $O^{o_l^h, o_l^w}$  (b) Processing of $O^{o_l^h, o_l^w+1}$

Figure 4: Illustration of the process described in Eq.(5) using $\xi^h = 2$ and processing two consecutive patches of $O$. Squares coloured with cyan signify the elements participating at the processing of $O^{o_l^h, o_l^w}$, and coloured with grey are the elements of $O^{o_l^h, o_l^w+1}$.

Next, $Q_j$ is input into ReLU activation function after batch normalization (BN). The process is described as below:

$O_j = ReLU\left(BN(Q_j)\right)$  (6)

Where, ReLU is non-linear activation function of linear rectification; BN is branch normalization; $O_j \in \mathbb{R}^{O_j^c, O_j^h, O_j^w}$ is output of $ReLU$. The output of final time-dilated convolution block $O_J \in \mathbb{R}^{O_J^c, O_J^h, O_J^w}$ is compressed into a one-dimensional vector $O' \in \mathbb{R}^{O_J^c \times O_J^h \times O_J^w}$, and then $O'$ is input into the subsequent classifier $Cls(\cdot)$. The classification recognition for underwater acoustic target is conducted as below:

$$\hat{Y} = Cls(O'). \quad (7)$$

Where, $\hat{Y} = [\hat{y}_1, \cdots, \hat{y}_C]$ is classification result predicted by the model for underwater acoustic target. This paper replaces RNN with time-dilated convolution network which can effectively implement long-term contextual, intra-class and inter-class activity modeling for UATR. The model parameters are optimized through minimizing the cross-entropy loss between $\hat{Y}$ and $Y$.

## 4. Model evaluation

In order to evaluate the proposed method, real data samples of civil ships are used, and this paper uses F1 score and accuracy (ACC) as indicators. Besides, this paper compares artificially designed features with those features extracted automatically with the proposed ATCNN. These artificially designed features include waveform, wavelet, MFCC, HHT, Mel frequency, non-linear auditory feature, spectrum and cepstrum. In addition, the histogram and (t-distributed stochastic neighbor embedding) t-SNE[54] are used for visualization for clustering performance of the proposed method. The proposed model is evaluated based on MXNET deep learning framework.

### 4.1 Dataset and data pre-processing

The experimental data is from real data samples of civil ships. The dataset contains small ship, big ship, and ferry. The data is sampled at anchorage ground, with frequency 48000Hz. In the experiment, 80% samples of each class are used as training set, while the rest 20% samples are used as testing set. Figure 5 shows time domain diagram of underwater noise. Figure 6 shows frequency domain diagram of underwater noise. The color level represents energy, with microvolt as the unit.

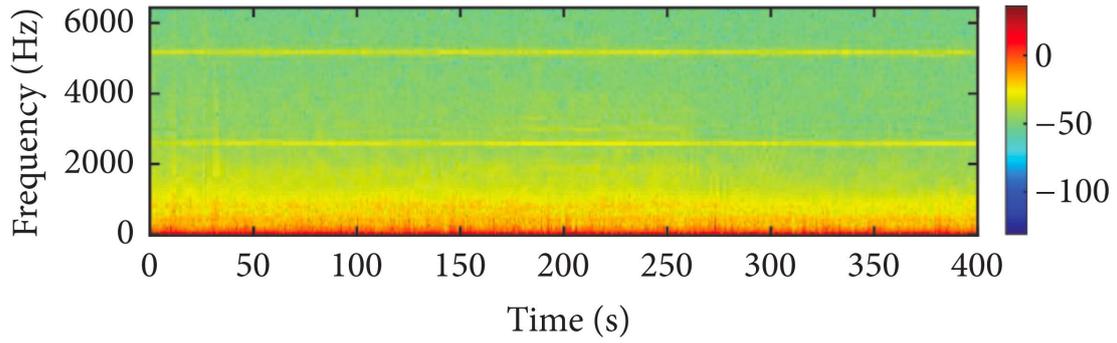

Figure 6: Frequency domain diagram of underwater noise.

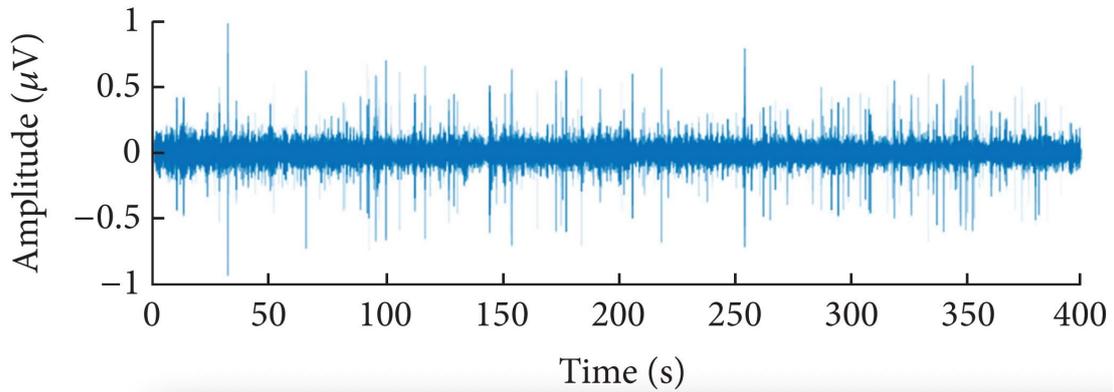

Figure 5: Time domain diagram of underwater noise.

Each record is generated as WAV audio file. The records include training dataset and testing dataset, 80% samples of each class are used as training set and the rest 20% samples of each class are used as testing set. Each record is divided into 10s audio segment, while training samples and testing samples are 45ms jumping 12.5ms. Because the network training and test are conducted on original time-domain data, any pre-processing is not required. Table 1 shows the total time and number of samples of each class in the training dataset and testing dataset.

Table 1. Experimental data description.

| Data Set | Class | No. Segments | Total Time (Hour) | No. Samples | Percentage |
| --- | --- | --- | --- | --- | --- |
| Training | small ship | 326 | 0.91 | 260736 | 25.9% |
|  | ferry | 560 | 1.55 | 447744 | 44.6% |
|  | big ship | 119 | 0.33 | 95424 | 9.5% |
| Test | small ship | 81 | 0.23 | 65184 | 6.5% |

| | ferry | 140 | 0.39 | 111936 | 11.1% |
|---|---|---|---|---|---|
| | big ship | 30 | 0.08 | 23856 | 2.4% |

It can be seen from Table 1 that the sample number of each class varies greatly; the sample number of small ships is 55.7% of the total sample number, more than half of total sample number; the sample number of big ships is only 11.9%. According to sample distribution of each class, the sample number of each class varies greatly. For different classification sample set, F1 score is a better indicator than accuracy. Therefore, it is required to evaluate the proposed model and investigate performance of the classification recognition with multiple indicators instead of one indicator. This paper uses F1 score and accuracy as indicators for evaluation.

**4.2 Hyper-parameters, indicators, and evaluation process**

In order to evaluate the proposed method, this paper uses the dataset of real time-domain radiated noise with sampling frequency 48000Hz, including three classes (C=3), i.e. "small ship", "big ship", and "ferry". This paper inputs 10s noise segments and divides them into 800 (T=800) vector sequences X with length 2176 (N=2176) through Hamming window function with 75% overlaps between continuous windows with length 45ms. Moreover, this paper normalizes all input vectors of X into zero to obtain the mean and unit variance; and then inputs X into learnable feature extractor which outputs radiated noise feature inspired by auditory perception. The length of feature vector is F=100.

Table 2 shows the architecture of learnable feature extractor. The learnable feature extractor is built based on above-mentioned one-dimensional DWS convolution, on which the first layer is standard one-dimensional convolution. All convolutions are followed by BN and ReLU activation function. Figure 6 compares layer with standard convolution, BN and ReLU activation function with DWS convolution layer with depthwise convolution, 1×pointwise convolution, BN and ReLU activation function.

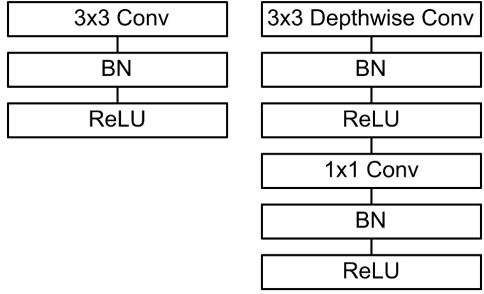

Figure 6. Left: Standard convolutional layer with BN and ReLU.

Right: Depthwise Separable convolutions with Depthwise and Pointwise layers followed by BN and ReLU.

The feature extractor network finally declines spatial resolution to 1 and considers depthwise and pointwise convolution as independent layers. There are five layers (L=5) in the feature extractor network. The proposed model structure makes more computations in 1×1 intensive convolution, which can be realized with highly optimized GEMM function. The general convolution is realized with GEMM, but an initial re-ranking called as "im2col" is required in memory to map it to GEMM. For example, this method is used in popular Caffe package [55]. 1×1 convolution can be realized directly with GEMM instead of re-ranking in memory. Therefore, GEMM is one of the optimal linear algebra algorithms. Among all DWS convolutions of the feature extractor network, 1×1 convolution takes 91% calculation time and uses 88% parameters, as shown in Table 3.

Table 2. Architecture of learnable feature extractor.

| Type | Stride | Filter Shape | Input Size |
| --- | --- | --- | --- |
| Conv1D | 50 | 204×1×64 dw | 2176×1 |
| Conv1D dw | 2 | 12×64 dw | 40×64 |
| Conv1D | 1 | 1×1×64×128 | 15×64 |
| Conv1D dw | 1 | 15×128 dw | 15×128 |
| Conv1D | 1 | 1×1×128×100 | 1×100 |

Table 3. Resource Per Depthwise Separable Layer.

| Type | Mult-Adds | Parameters |
| --- | --- | --- |
| Conv1D dw | 11520 | 832 |
| Conv1D | 122880 | 8320 |
| Conv1D dw | 1920 | 2048 |
| Conv1D | 12800 | 12900 |

The architecture of time-dilated convolutions inspired by the language model is shown in Table 4. The one-dimensional underwater acoustic feature vector $H' \in \mathbb{R}^F$ at continuous $T$ time output by the learnable feature extractor, two-dimensional matrix $I \in \mathbb{R}^{T \times F}$ combined by $T$ number of $H'$ (where, $T = 800$ & $F = 100$), and $I$ are input into the time-dilated convolution network inspired by language model. The time-dilated convolution dilates only on time dimension and this paper uses the dilation factor $\xi^h = 12$. All convolution layers are followed by BN and ReLU activation function, but pooling layer is not provided with non-linear activation function. The final pooling layer is input into softmax layer for classification. There are five layers (J=5) in time-dilated convolution network.

Table 4. Architecture of time-dilated convolutions.

| Type | Stride | Dilation | Filter Shape | Input Size |
|---|---|---|---|---|
| Conv2D Dilation | 1×1 | 12×1 | 3×3×1×64 | 800×100×1 |
| Max Pool | 2×2 | 1×1 | Pool 2×2 | 776×98×64 |
| Conv2D Dilation | 1×1 | 12×1 | 3×3×64×128 | 388×49×64 |
| Max Pool | 2×2 | 1×1 | Pool 2×2 | 364×47×128 |
| Conv2D Dilation | 1×1 | 12×1 | 3×3×128×256 | 182×23×128 |
| Avg Pool | 2×2 | 1×1 | Pool 2×2 | 158×21×256 |
| Conv2D Dilation | 1×1 | 12×1 | 3×3×256×512 | 79×10×256 |
| Avg Pool | 2×2 | 1×1 | Pool 2×2 | 55×8×512 |
| Conv2D Dilation | 1×1 | 12×1 | 3×3×512×512 | 27×4×512 |
| Avg Pool | 2×2 | 1×1 | Pool 2×2 | 3×2×512 |
| Softmax | | | Classifier | 1×1×3 |

In order to evaluate performance of the proposed method, this paper uses F1 score (the larger the better) and accuracy (ACC, the larger the better) as indicators for evaluation. This paper compares recognition model of artificially designed features, one-dimensional depthwise convolution network [22] without time-dilated convolution, and the proposed ATCNN. These artificially designed features include waveform, wavelet, MFCC, HHT, Mel frequency, non-linear auditory feature, spectrum and cepstrum. All classifiers here are SVMs. ATCNN is evaluated on deep learning architecture MXNET[56]; MXNET Python library runs on nvidia RTX graphic card; and the asynchronous gradient similar to Inception V3[52] is used to decline optimizer RMSProp[57]. Table 5 shows hyper-parameters of the proposed model.

Table 5. Hyper-parameters of the proposed model.

| Parameters | Values |
|---|---|
| **Learning Rate** | **0.001** |
| **Batchsize** | **800** |
| **Epochs** | **100** |
| **Optimizer** | **RMSprop** |

Some researchers of neurosciences found that the brain can change its structure and functions to meet learning demands [40]. With the drive of different audio stimulation, frequency related tissues such as auditory cortex and auditory midbrain can adjust the receptive field of frequency to complete auditory task better [41,42]. In the model, ATCNN realizes frequency perception and decomposition of the auditory system to a certain extent. Unlike large-scale model training, this paper uses few regularization and data processing technologies, because it is not easy to conduct overfitting for small-scale model. With the drive of time-domain signals of ship-radiated noise, all parameters of ATCNN are learnt from real data. Moreover, it is possible to learn and adjust frequency decomposition and perception capacity of the ATCNN. The shaping of such frequency perception and decomposition can reflect brain shaping.

## 5. Results and discussion

Based on MXNET deep learning framework, this experiment uses Python language to write related codes; and configures the server running neural network as below: 64-bit Ubuntu 16.04 OS,

64 GB memory, 52 CPU kernels and TITAN RTX GPU accelerated computing card from NVIDIA company.

This paper uses original time-domain ship-radiated noise data for model training and testing. The training parameters include epochs of 100 times, training batch of 800, and training rate of 0.9. The detailed training process is shown in Figure 7.

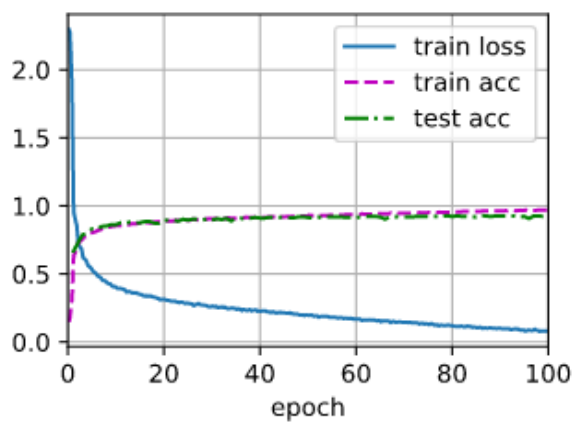

Figure 7. The training process of the model

As shown in Figure 7, there is no fitting and underfitting, and gradient vanishing or exploding problem during model training. The model trained with real measurement data obtains 95.9% recognition accuracy for training data and 90.9% recognition accuracy for testing data, reflecting high recognition accuracy.

With 90.9% recognition accuracy, the model works well for underwater acoustic data with strong noise. Because sample number of each class varies greatly, this paper provides confusion matrix for recognition result of the proposed model, as shown in Figure 8. The confusion matrix describes more details, where each line represents real label and each row represents predicted label.

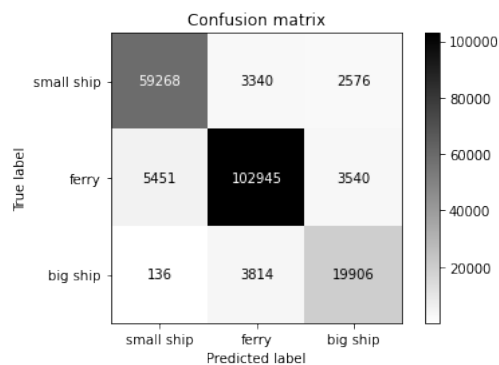

Figure 8．Confusion matrix of the proposed model obtained from testing data.

As shown in Figure 8, recognition result is stable among classes, indicating great stability in recognition.

For unbalanced dataset in multi-classification problem, the most common indicator is the multi-classification F1 score. A binary classification score is calculated for each class and this class is positive, while all other classes are negative. Then, one of the following strategies is used to average these F1 scores. "Macro average" is to calculate unweighted F1 score. It gives same weight for all classes despite their sample sizes. The classification report provides this value. "Micro average" is to calculate total number of false positive, false negative, and true positive, and then calculate accuracy, recall, and F1 score with these counts. If you treat each sample equivalently, then "micro average" score is recommended; if you treat each class equivalently, then "macro average" is recommended. This paper calculates accuracy, recall, and F1 score of each class, as shown in Table 6:

Table 6. Precision, Recall, and F1 score for each class

| Class | Precision | Recall | F1 score | Support |
| --- | --- | --- | --- | --- |
| small ship | 0.91 | 0.91 | 0.91 | 65184 |
| ferry | 0.94 | 0.92 | 0.93 | 111936 |
| big ship | 0.76 | 0.83 | 0.80 | 23856 |
| Accuracy | | | 0.91 | 6.5% |
| Macro avg | 0.87 | 0.89 | 0.88 | 11.1% |
| Weighted | 0.91 | 0.91 | 0.91 | 2.4% |

As shown in Table 6, the model has a good F1 score, indicating great classification accuracy and stability. The small ship and ferry obtain the best results with F1 score of 0.91 and 0.93. The big ship obtains the worst results with accuracy of 0.76, recall of 0.83, and F1 score of 0.80. The reason may be that small ship has similar mechanical system with ferry, or some small ships are passing through during sampling for ferry.

In order to simulate practical application of recognition for ship-radiated noise, the classification accuracy of each acoustic event is used to measure classification performance of the model. It is defined as percentage of acoustic event with correct classification in all acoustic events. The accuracy values of the proposed model and compared models are shown in Table 7.

Table 7. Classification results of proposed model and compared models.

| Input | Methods | Accuracy |
| --- | --- | --- |
| HOS[59] | SVM | 85.1% |
| Waveform [60,8] | SVM | 78.9% |
| Wavelet [10] | SVM | 84.3% |
| MFCC [62] | SVM | 79.1% |
| Mel-frequency | SVM | 84.6 |
| Nolinear auditory | SVM | 86.7 |
| Spectral [63,64] | DNN [58] | 87.0% |
| Cepstral [9,61] | DNN [58] | 86.9% |
| Raw time domain data | CNN model[22] | 88.4% |
| Raw time domain data | CRNN model[25] | 89.2% |
| Raw time domain data | ATCNN | 90.1% |

As shown in above Table, compared with traditional UATR method, the proposed model effectively improves classification accuracy of UATR.

Next, feature extraction is analyzed for the model. Here, a histogram is created for each feature to calculate frequency of data point with certain features in a specific range (called as "bin"). Each Figure contains two histograms, from which distribution of each feature in each class can be seen and those features to be distinguished greatly can be seen. Most parts of the histogram are overlapped if there is no information content. However, it seems that some features have a relatively large information content, because the intersection in histogram is small. This paper extracts 100-

dimension features for underwater acoustic target. Figure 9 shows some histograms of features extracted by the model.

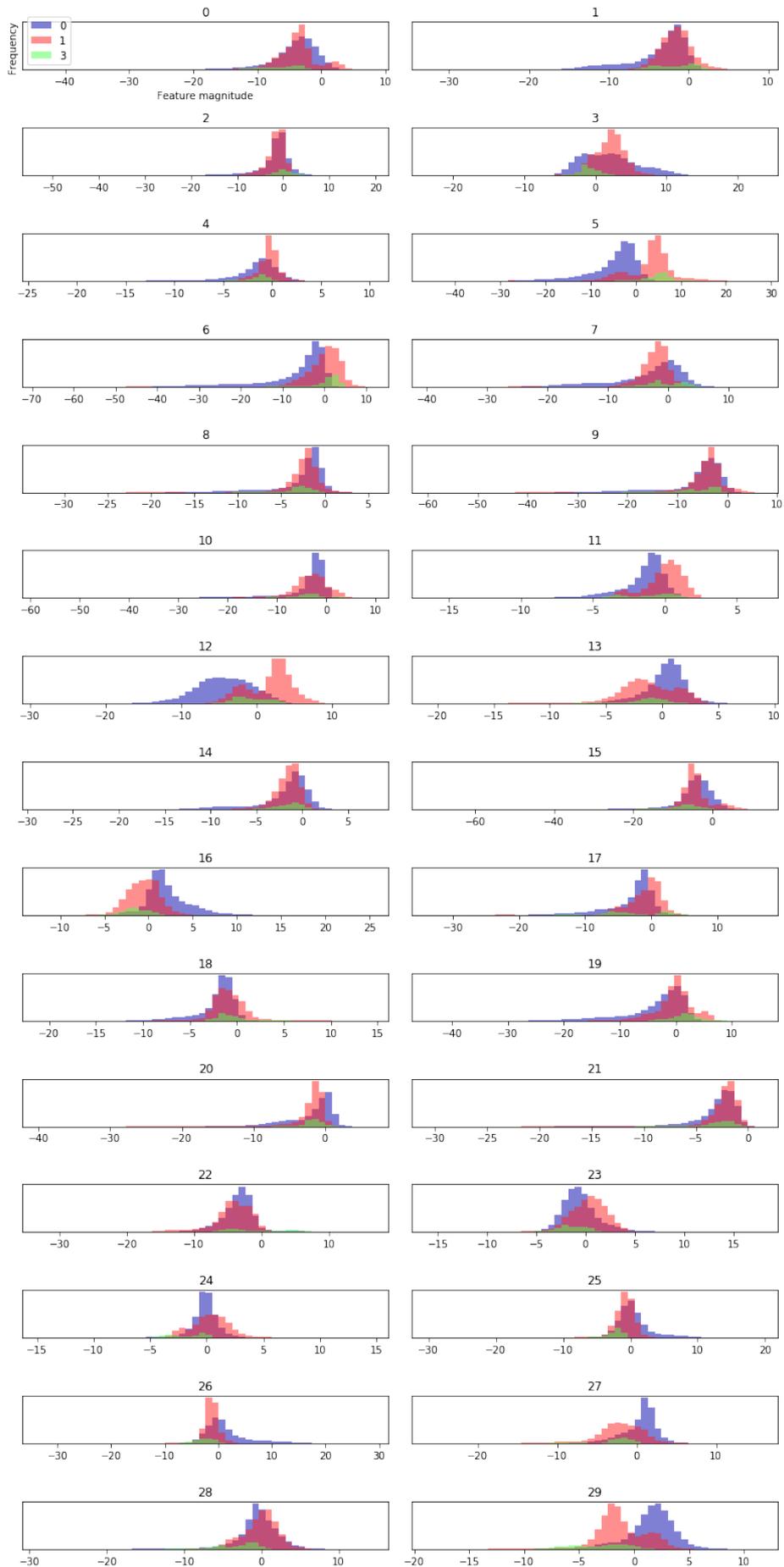

Figure 9. Histogram of features learned from the proposed model

0 is small ship; 1 is big ship; 2 is ferry

As shown in Figure 9, there exist some features with relatively more information content and significant distinction for each class. This indicates effectiveness of the feature extraction. The manifold learning algorithm allows more complex mapping and usually enables better visualization. There, t-SNE algorithm is particularly useful. It is used for feature visualization of underwater acoustic target. Figure 6 shows the scatter plot of 100 features.

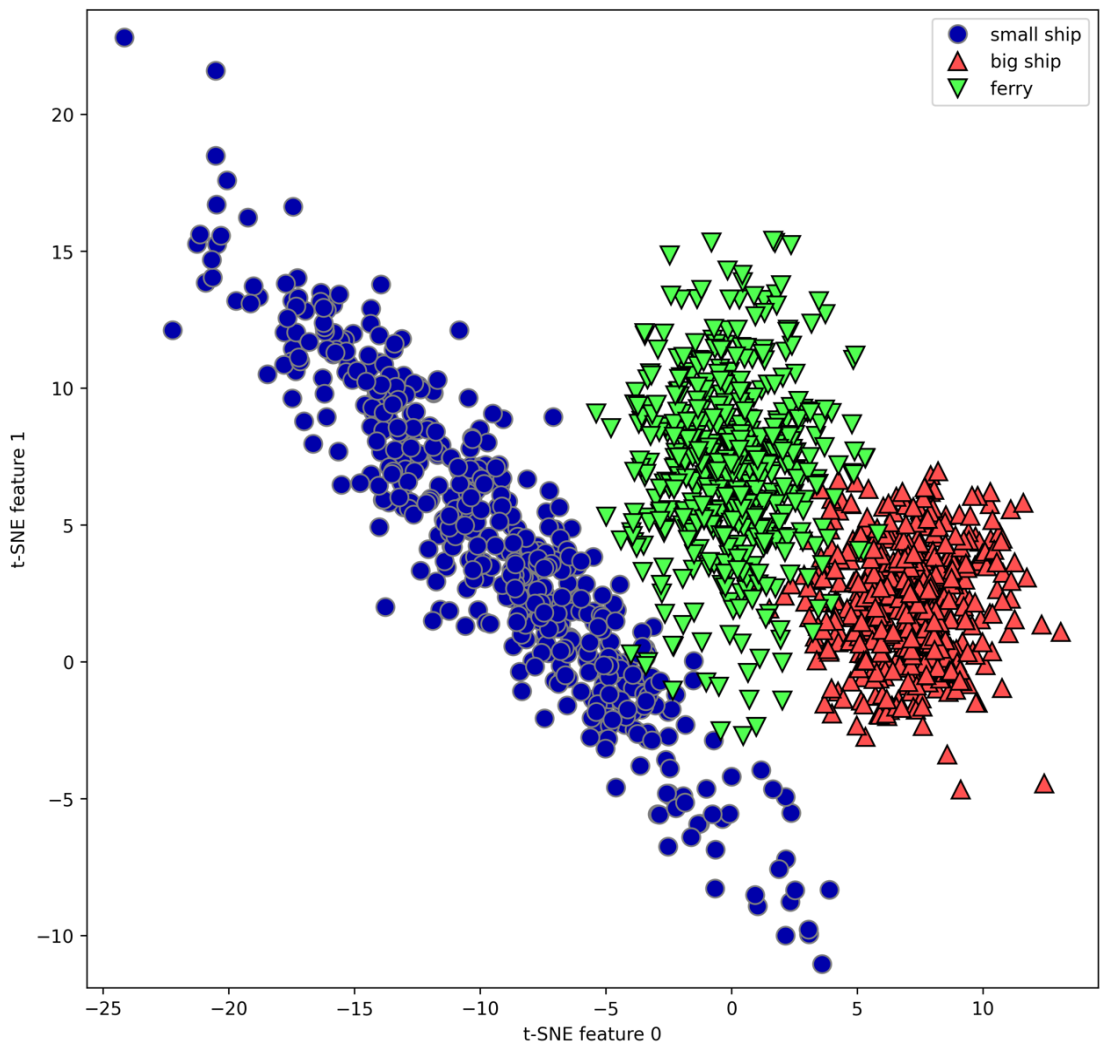

Figure 10. Scatter plot of features learned from proposed model using two components found by t-SNE

As shown in Figure 10, the corresponding number of each class is used here as a symbol to indicate position of each class. It can be seen that each class is separated properly. This indicates that features extracted by the model have a great performance in clustering, and demonstrated through visualization that features extracted for underwater acoustic target are greatly separable and stable. This paper creates a feature dataset for underwater acoustic target with obtained feature vectors, and uses support vector for classification recognition for the dataset to obtain recognition an accuracy of 90.1%. This indicates that recognition accuracy of the model mainly depends on feature extraction and great feature engineering provides convenience for design of classifier. Moreover, SVM is an excellent non-linear classifier and can reach higher classification accuracy compared with the linear classifier.

## 6. Conclusions

This paper proposes an ATCNN which recognizes ship-radiated noise from original time-domain waveform in end-to-end mode. This paper uses DWS convolution network to extract deep features of information containing objectives and reflect the structure of extracting deep acoustic information in auditory system. Through the convolutional decomposition for different frequency components of the ship-radiated noise, frequency distribution of the ship-radiated noise is revealed. The time-dilated convolution is used for long-term contextual modeling, thus intra-class and inter-class information can be fully used for UATR like the language model. Moreover, inspired by plastic neural mechanism, this paper learns and optimizes all parameters in ATCNN model under the drive of time-domain ship-radiated noise to complete the UATR task. The deep features of underwater acoustic target extracted with the proposed method here are greatly separable and stable. When they are tested on the dataset of real acoustic signals of civil ships, the average recognition rate for classification reaches 90.9%. Although the recognition rate is high, it has a certain gap with real-world application and therefore shall be further improved. The experimental results also indicate that extracted 100-dimension features of underwater acoustic target are greatly separable and stable. Therefore, the deep learning method based on auditory perception has great potential to improve classification performance of UATR.

**Acknowledgement**
This work was supported by the Basic Scientific Research project (No. JCKY2017207B042).

**Reference**

[1] Xu Ji, Huang Zhaoqiong, Li Chen, et al. Advances in Underwater Target Passive Recognition Using Deep Learning[J]. Journal of Signal Processing, 2019, 35(9): 1460-1475. DOI: 10.16798/j.issn.1003-0530.2019.09.003.
[2] Lourens J G. Classification of ships using underwater radiated noise. Conference on Communications & Signal Processing,1988,161:130-134.
[3] Rajagopal R, Sankaranarayanan B, Rao P R. Target classification in a passive sonar-An expert system approach. IEEE ICASSP-90, 2002, 5(5): 2911-2914.
[4] Maksym J N, Bonner A J, Dent C A, et al. Machine analysis of acoustical signal. Pattern Recognition,1983,16(6):615-625.
[5] Amab D, Arun K, Rajendar B. Feature analyses for marine vessel classification using passive sonar. UDT, 2005.21-23.
[6] Farrokhrooz M, Karim M. Ship noise classification using probabilistic neural network and AR model coefficients . Oceans , 2005 , 2 : 1107-1110 .
[7] Nii H P, Feigenbaum E, Anton J, et al. Signal-to-symbol transformation: Reasoning in the HASP/SIAP program. IEEE International Conference on Acoustics, Speech, & Signal Processing,1984,9:158-161.
[8] Meng, Q.; Yang, S. A wave structure based method for recognition of marine acoustic target signals. J. Acoust. Soc. Am. 2015, 137, 2242–2242, doi:10.1121/1.4920186.
[9] Das, A.; Kumar, A.; Bahl, R. Marine vessel classifification based on passive sonar data: The cepstrum-based approach. IET Radar Sonar Navig. 2013, 7, 87–93, doi:10.1049/iet-rsn.2011.0142.
[10] Wei, X.; Li, G.; Wang, Z.Q. Underwater target recognition based on wavelet packet and principal component analysis. Comput. Simul. 2011, 28, 8–290, doi:10.1090/S0002-9939-2011-10775-5.
[11] Yang, H.; Gan, A.; Chen, H.; Yue, P.; Tang, J.; Li, J. Underwater acoustic target recognition using SVM ensemble via weighted sample and feature selection. In Proceedings of the 13th International Bhurban Conference on Applied Sciences and Technology, Islamabad, Pakistan, 12–16 January 2016, doi:10.1109/IBCAST.2016.7429928.
[12] Filho, W.S.; Seixas, J.M.D.; Moura, N.N.D. Preprocessing passive sonar signals for neural classification. Iet Radar Sonar Navig. 2011, 5, 605–612, doi:10.1049/iet-rsn.2010.0157.
[13] Sainath, T.N.; Kingsbury, B.; Mohamed, A.R.; Ramabhadran, B. Learning filter banks within a deep neural network framework. In Proceedings of the 2013 IEEE Workshop on Automatic Speech Recognition and Understanding (ASRU), Olomouc, Czech Republic, 8–12 December 2013; IEEE: Piscataway, NJ, USA, 2014; pp. 297–302.
[14] Hinton G, Salakhutdinov R. Reducing the dimensionality of data with neural networks. Science,2006,313(5786):504-507.
[15] Hinton G E, Osindero S, Teh Y W. A fast learning algorithm for deep belief nets. Neural Computation,2006,18(7):1527-1554.
[16] Weinberger, N.M. Experience-dependent response plasticity in the auditory cortex: Issues, characteristics, mechanisms, and functions. In Springer Handbook of Auditory Research; Springer: Cham, Switzerland, 2004; doi:10.1007/978-1-4757-4219-0_5.
[17] I. Sutskever, O. Vinyals, and Q. V. Le, "Sequence to sequence learning with neural networks," in Advances in Neural Information Processing Systems 27, Z. Ghahramani, M. Welling, C. Cortes, N. D. Lawrence, and K. Q. Weinberger, Eds. Curran Associates, Inc., 2014,

pp. 3104–3112. [Online]. Available: http://papers.nips.cc/paper/ 5346-sequence-to-sequence-learning-with-neural-networks.pdf

[18] M. Auli, M. Galley, C. Quirk, and G. Zweig, "Joint language and translation modeling with recurrent neural networks," in Proceedings of the 2013 Conference on Empirical Methods in Natural Language Processing. Seattle, Washington, USA: Association for Computational Linguistics, Oct. 2013, pp. 1044–1054. [Online]. Available: https://www.aclweb.org/anthology/D13-1106

[19] S. Bengio, O. Vinyals, N. Jaitly, and N. Shazeer, "Scheduled sampling for sequence prediction with recurrent neural networks," in Proceedings of the 28th Interna tional Conference on Neural Information Processing Systems - Volume 1, ser. NIPS'15. Cambridge, MA, USA: MIT Press, 2015, pp. 1171–1179. [Online]. Available: http://dl.acm.org/citation.cfm?id=2969239.2969370

[20] K. Drossos, S. Gharib, P. Magron, and T. Virtanen, "Language modelling for sound event detection with teacher forcing and scheduled sampling," in Workshop on Detection and Classifification of Acoustic Scenes and Events (DCASE), Oct. 2019.

[21] Cheng Jinsheng, Du Xuanmin, Zeng Sai. Research on Audio Feature Extraction and Recognition of Underwater Targets Using Deep Learning Method[C]//Chinese A-coustic Society. Proceedings of the National Acoustics Conference in 2018,2018:2.(in Chinese)

[22] Gang Hu, et al. "Deep Learning Methods for Underwater Target Feature Extraction and Recognition." Computational Intelligence and Neuroscience, 2018, (2018-3-27) 2018(2018):1-10.

[23] Lang Zeyu. Research on underwater target feature extraction based on convolutional neural network[D]. Harbin: Harbin Engineering University, 2017.(in Chinese)

[24] Wang Nianbin, He Ming, Wang Hongbin, et al. Fast dimensionality reduction convolution model for underwater target recognition[J/OL]. Journal of Harbin Engineering University:1-6[2019-05-05].(in Chinese)

[25] Lu Anan. Underwater Acoustic Classification Based on Deep Learning[D]. Harbin: Harbin Engineering University, 2017.(in Chinese)

[26] Zeng Sai, Cheng Jinsheng. Application of Heterogeneous Multimodal Deep Learning Method in Underwater Target Recognition [ C ] // Chinese Acoustic Society , 2018 : 2(in Chinese)

[27] Zhang Shaokang, Tian Deyan. Intelligent classification method of Mel frequency cepstrum coefficient for under-water acoustic targets [J/OL]. Journal of Applied Acous-tics,2019(2):267-272.(in Chinese)

[28] K. Drossos, S.I. Mimilakis, S. Gharib. Y. Li, and T. Virtanen, "Sound event detection with depthwise separable and dilated convolutions" in 2020 International Joint Conference on Neural Networks(IJCNN ), Jul.2020.

[29] E. Fonseca, M. Plakal, F. Font, D. P. W. Ellis, and X. Serra"Audio tagging with noisy labels and minimal supervision" in Workshop on Detection and Classification of Acoustic Scenes and Events(DCASE), Oct.2019.

[30] Howard A G , Zhu M , Chen B , et al. MobileNets: Efficient Convolutional Neural Networks for Mobile Vision Applications[J]. 2017.

[31] L. Sifre. "Rigid-motion scattering for image classification" Ph.D. dissertation, Ecole Polytechnique, CMAP,2014.

[32] F. Yu and V. Koltun, "Multi-scale context aggregation by dilated convolutions" in International Conference on Learning Representations (ICLR), May2016.

[33] M. Holschneider, R. Kronland-Martinet, J.Morlet,P. Tchamitchian, "A real-time algorithm for signal analysis with the help of the wavelet transform" in Wavelets, J.-M.Combes, A. Grossmann, and P. Tchamitchian, Eds. Berlin Heidelberg: Springer Berlin Heidelberg, 1990, pp. 286-297.

[34] M. J. Shensa, "The discrete wavelet transform:wedding a trous and mallat algorithms" IEEE Transactions on Signal Processing, vol. 40, no. 10, pp. 2464-2482, Oct. 1992.

[35] Y. Li, M. Liu, K. Drossos, and T. Virtanen"Sound event detection via dilated convolutional recurrent neural networks " in ICASSP 2020 - 2020 IEEE International Conference on Acoustics, Speech and Signal Processing (ICASSP), 2020, pp. 286-290.

[36] Gazzaniga, M.; Ivry, R.B.; Mangun, G.R. Cognitive Neuroscience The Biology of the Mind; W. W. Norton & Company: New York, NY, USA, 2018.

[37] Dallos, P.; Popper, A.N.; Fay, R.R. The Cochlea. In Springer Handbook of Auditory Research; Springer: Cham, Switzerland, 1996; Volume 65, pp. 291–312, doi:10.1007/978-1-4612-0757-3.

[38] Brundin, L.; Flock, A.; Canlon, B. Sound-induced motility of isolated cochlear outer hair cells is frequency-specific. Nature 1989, 342, 814, doi:10.1038/342814a0.

[39] Schreiner, C.E.; Langner, G. Laminar fine structure of frequency organization in auditory midbrain. Nature 1997, 388, 383–386, doi:10.1038/41106.

[40] Schreiner, C.E.; Read, H.; Sutter, M. Modular organization of frequency integration in primary auditory cortex. Ann. Rev. Neurosci. 2000, 23, 501–529, doi:10.1146/annurev.neuro.23.1.501.

[41] Strainer, J.C.; Ulmer, J.L.; Yetkin, F.Z.; Haughton, V.M.; Daniels, D.L.; Millen, S.J. Functional MR of the primary auditory cortex: An analysis of pure tone activation and tone discrimination. AJNR Am. J. Neuroradiol. 1997, 18, 601–610.

[42] Talavage, T.M.; Ledden, P.J.; Benson, R.R.; Rosen, B.R.; Melcher, J.R. Frequency-dependent responses exhibited by multiple regions in human auditory cortex. Hear. Res. 2000, 150, 225–244, doi:10.1016/S0378-5955(00)00203-3.

[43] Kolb, B.; E., R.G.T. Brain plasticity and behavior. Curr. Dir. Psychol. Sci. 2003, 12, 1–5, doi:10.2307/20182820.

[44] Robertson, D.; Irvine, D.R.F. Plasticity of frequency organization in auditory cortex of guinea pigs with partial unilateral deafness. J. Comp. Neurol. 2010, 282, 456–471, doi:10.1002/cne.902820311.

[45] Weinberger, N.M. Learning-induced changes of auditory receptive fields. Curr. Opin. Neurobiol. 1993, 3, 570, doi:10.1016/0959-4388(93)90058-7.

[46] Zhang Yufan, Zhang Jing. Understanding convolution from filter [J]. Electronic production, 2019, 000 (011): 46-47, 50.(in Chinese)

[47] Moore, B.C.J. Cochlear Hearing Loss: Physiological, Psychological and Technical Issues, 2nd ed.; John Wiley and Sons: Hoboken, NJ, USA, 2008.

[48] Gelf, S.A. Hearing: An Introduction to Psychological and Physiological Acoustics; CRC Press: Boca Raton, FL, USA, 2009; p. 320.

[49] Lecun, Y.; Bottou, L.; Bengio, Y.; Haffner, P. Gradient-based learning applied to document recognition. Proc. IEEE 1998, 86, 2278–2324, doi:10.1109/5.726791.

[50] Lecun, Y.; Bengio, Y.; Hinton, G.E. Deep learning. Nature 2015, 521, 436, doi:10.1038/nature14539.

[51] F. Chollet, "Xception: Deep learning with depthwise separable convolutions," in 2017 IEEE Conference on Computer Vision and Pattern Recognition (CVPR), Jul. 2017, pp. 1800–1807.

[52] C. Szegedy, V. Vanhoucke, S. Ioffffe, J. Shlens, and Z. Wojna, "Rethinking the inception architecture for computer vision," in 2016 IEEE Conference on Computer Vision and Pattern Recognition (CVPR), Jun. 2016, pp. 2818–2826.

[53] S. Ioffe and C. Szegedy, "Batch normalization: Accelerating deep network training by reducing internal covariate shift," in ICML, 2015.

[54] Hinton, G.E. Visualizing high-dimensional data using t-SNE. Vigiliae Christianae 2008, 9, 2579–2605.

[55] Y. Jia, E. Shelhamer, J. Donahue, S. Karayev, J. Long, R. Girshick, S. Guadarrama, and T. Darrell. Caffe: Convolutional architecture for fast feature embedding. arXiv preprint arXiv:1408.5093, 2014.

[56] Tianqi Chen, Mu Li, et al. MXNet: A Flexible and Effificient Machine Learning Library for Heterogeneous Distributed Systems, 2015. Software available from tmxnet.apache.org, 12, 2015.

[57] T. Tieleman and G. Hinton. Lecture 6.5-rmsprop: Divide the gradient by a running average of its recent magnitude. COURSERA: Neural Networks for Machine Learning, 4(2), 2012.

[58]. Yue, H.; Zhang, L.; Wang, D.; Wang, Y.; Lu, Z. The classification of underwater acoustic targets based on deep learning methods. Adv. Intell. Syst. Res. 2017, 134, 526–529.

[59] Chen Fenglin, Lin Zhengqing, Peng Yuan, et al. High order statistics feature extraction and feature compression of ship radiated noise [D], 2010 (in Chinese)

[60] Meng, Q.; Yang, S.; Piao, S. The classification of underwater acoustic target signals based on wave structure and support vector machine. J. Acoust. Soc. Am. 2014, 136, 2265.

[61] Santos-Domínguez, D.; Torres-Guijarro, S.; Cardenal-López, A.; Pena-Gimenez, A. ShipsEar: An underwater vessel noise database. Appl. Acoust. 2016, 113, 64–69.

[62] Zhang, L.; Wu, D.; Han, X.; Zhu, Z. Feature extraction of underwater target signal using Mel frequency
cepstrum coeffificients based on acoustic vector sensor. J. Sens. 2016, 2016, 7864213.

[63] Kamal, S.; Mohammed, S.K.; Pillai, P.R.S.; Supriya, M.H. Deep learning architectures for underwater target recognition. In Proceedings of the 2013 International Symposium on Ocean Electronics, Kochi, India, 23–25 October 2013; IEEE: Piscataway, NJ, USA, 2014; pp. 48–54.

[64]18. Cao, X.; Zhang, X.; Yu, Y.; Niu, L. Deep learning-based recognition of underwater target. In Proceedings of the IEEE International Conference on Digital Signal Processing, Beijing, China, 16–18 October 2017; IEEE: Piscataway, NJ, USA, 2018; pp. 89–93.